\documentstyle[12pt,epsfig]{article} 
\begin{document}

\begin{center}
{\large {\bf YAG:Nd crystals as possible detector to search for
$2\beta$ and $\alpha$ decay of neodymium}}

\vskip 0.2cm

F.A.~Danevich\footnote{Corresponding author. {\it E-mail address:}
danevich@kinr.kiev.ua}, V.V.~Kobychev, S.S.~Nagorny, V.I.~Tretyak

\vskip 0.2cm

\noindent{\it Institute for Nuclear Research, MSP 03680 Kiev,
Ukraine}

\end{center}

\hspace{1.0in}
\vskip 0.5cm
\begin{abstract}

Energy resolution, $\alpha/\beta$ ratio, pulse-shape
discrimination for $\gamma$ rays and $\alpha$ particles,
radioactive contamination were studied with neodymium doped
yttrium-aluminum garnet (YAG:Nd). Applicability of YAG:Nd
scintillators to search for $2\beta$ decay and $\alpha$ activity
of natural neodymium isotopes are discussed.
\end{abstract}

\noindent

\vskip 0.4cm


\vskip 0.4cm


\vskip 1.0cm

\section{INTRODUCTION}

The great interest to the double beta ($2\beta$) decay search
is related, in particular, with the recent evidence of neutrino
oscillations which strongly suggests that neutrinos have nonzero mass
\cite{Ver02,Futur,Ell02,DBD-tab}. While oscillation experiments
are sensitive to the neutrinos mass difference, only the measured
neutrinoless ($0\nu$) double beta decay rate could set the
Majorana nature of the neutrinos and give the absolute scale of
the effective neutrino mass.

$^{150}$Nd is one of the most promising candidate for $2\beta$
decay study due to its large transition energy ($Q_{2\beta}$). As
a result, the calculated value of the phase space integral
$G_{mm}^{0\nu}$ of the $0\nu2\beta$ decay of $^{150}$Nd is the
largest one among 35 possible $2\beta^-$ decay candidates
\cite{Doi85,Suh98}. The theoretical predictions of the product
$T_{1/2}^{0\nu}\cdot\langle m_\nu\rangle^2$ are in the interval
$3.4\times 10^{22}-3.4\times 10^{24}$ yr (half-lives for the
neutrino mass mechanism with $\langle m_{\nu} \rangle = 1$ eV)
\cite{NME}. Moreover,
the larger the $Q_{2\beta}$ energy, the simpler, from an
experimental point of view, is to overcome background problems.
Note that the background from natural radioactivity drops sharply
above 2615 keV, which is the energy of the $\gamma$ from
$^{208}$Tl decay ($^{232}$Th family). In addition, a contribution of
cosmogenic activation, which is important problem of the next
generation $2\beta$ decay experiments \cite{CARVEL}, decreases at higher
energies.

There exist no appropriate detector containing neodymium, which
could serve as both source and detector simultaneously (so called
"active" source experimental method \cite{Futur}). It should
be noted that this experimental approach will assure high registration
efficiency which is especially important if very expensive
enriched $^{150}$Nd isotope would be used in an experiment.

The purpose of our work was investigation of scintillation
properties and preliminary check of radioactive contamination of
neodymium doped yttrium-aluminum garnet (YAG:Nd) as a possible
detector for double beta decay experiment with $^{150}$Nd. The
active source method allows also to investigate other rare processes in
Nd nuclei, such as $\alpha$ decay of naturally occurring neodymium
isotopes.

\section{MEASUREMENTS AND RESULTS}

\subsection{Energy resolution and $\alpha/\beta$ ratio}

Main properties of YAG:Nd (chemical formula
Y$_3$Al$_5$O$_{12}$:Nd) are presented in Table 1. These crystals
are well known and their production is well developed due to their
wide application in laser technique. The scintillation properties
of cerium doped yttrium-aluminum garnet (YAG:Ce) have been studied
in \cite{YAG:Ce}, properties of YAG:Yb have been reported in
\cite{Anto01}, whereas YAG:Nd, to our knowledge,
was never studied as a scintillator.

\begin{table}[tbp]
\caption{Properties of YAG:Nd crystal scintillators}
\begin{center}
\begin{tabular}{|l|l|}
\hline
Density (g/cm$^3$)                     & 4.56  \\
Melting point                          & 1970$^\circ$C \\
Crystal structure                      & Cubic Garnet \\
Hardness (Mohs)                        &  8.5   \\
% Wavelength of emission maximum (nm)  &  $\approx500$ ?  \\
Refractive index                       &  1.82   \\
Average decay time$^{\ast}$            &  4 $\mu$s \\
Photoelectron yield relatively to NaI(Tl)$^{\ast}$  & 8\% \\
\hline
\multicolumn{2}{l}{$^{\ast}$For $\gamma$ rays, at room temperature.} \\
\end{tabular}
\end{center}
\end{table}

All the measurements were carried out for the $\oslash 17 \times
6$ mm YAG:Nd crystal with $\approx$2 mol\% of Nd. The mass of the
crystal is 7.16 g. Photoelectron yield was estimated with the
Philips XP2412 bialkali photomultiplier (PMT) as 8\% of NaI(Tl).
The energy resolution FWHM=13.6\% was measured for 662 keV
$\gamma$ line of $^{137}$Cs with the YAG:Nd crystal wrapped by
PTFE reflector tape and optically coupled to the PMT XP2412. A
substantial improvement of the light output ($\approx$22\%) and
energy resolution was achieved by placing the crystal in liquid
(silicone oil with index of refraction $\approx$1.5). The crystal
was fixed in center of the teflon container $\oslash 70 \times 90$
mm and viewed by two PMTs XP2412. Fig. 1 demonstrates the energy
spectra of $^{137}$Cs and $^{207}$Bi obtained in such a way with
the YAG:Nd crystal. The energy resolution of 9.3\% was obtained
with the YAG:Nd crystal scintillator for 662 keV $\gamma$ line of
$^{137}$Cs.

\nopagebreak
\begin{figure}[ht]
\begin{center}
\mbox{\epsfig{figure=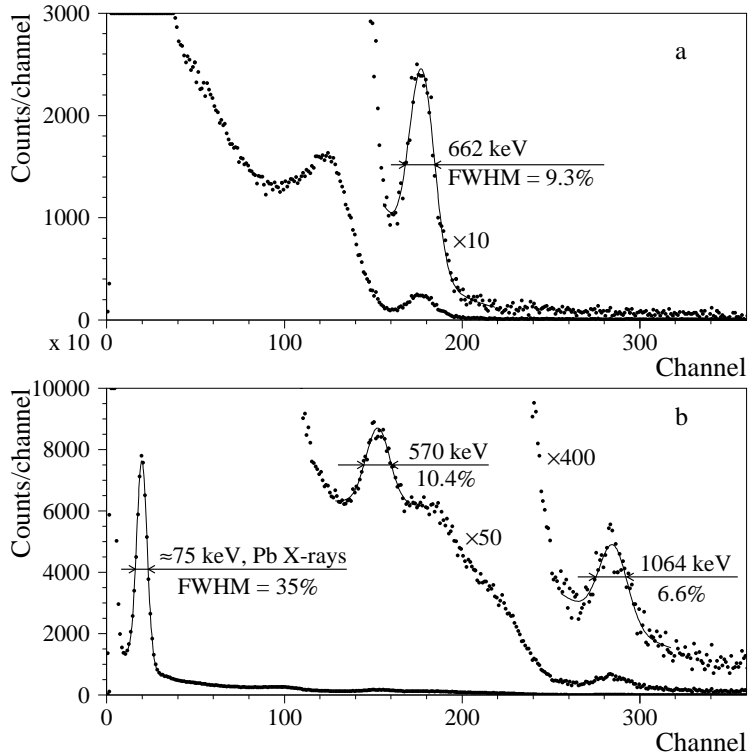,height=10.0cm}}
\caption {Energy spectra of $^{137}$Cs (a) and $^{207}$Bi
(b) $\gamma$ rays measured with the YAG:Nd scintillation crystal.
The crystal was located in liquid and viewed by two distant PMTs
(see text).}
\end{center}
\end{figure}

We estimate the energy resolution of the YAG:Nd scintillation
detector at the energy $Q_{2\beta}$ of $^{150}$Nd using the
results of measurements with $^{137}$Cs and $^{207}$Bi $\gamma$
sources. These data were fitted by function
FWHM(keV)$~=a+\sqrt{b\times E_{\gamma}}$ with values $a=2$ keV and
$b=5.2$ keV, where energy of $\gamma$ quanta $E_{\gamma}$ is in
keV. The energy resolution FWHM$\approx$4\% could be achieved at
the $Q_{2\beta}$ energy of $^{150}$Nd.

The $\alpha/\beta$ ratio was measured with the help of the
collimated $\alpha$ particles of a $^{241}$Am source. As it was
checked by surface-barrier detector, the energy of $\alpha$
particles was reduced to about 5.25 MeV by $\approx 1$ mm of air,
due to passing through the collimator. Fig. 2 shows the energy
spectrum of the $\alpha$ particles measured by the YAG:Nd
scintillator. The $\alpha/\beta$ ratio is 0.33. The YAG:Nd crystal
was irradiated in three perpendicular directions with aim to check
a possible dependence of the $\alpha $ signal on direction of
irradiation. While earlier such a dependence has been found for
CdWO$_4$ \cite{W-alpha} and ZnWO$_4$ \cite{ZWO} scintillators, we
did not observe it for YAG:Nd crystal.

\nopagebreak
\begin{figure}[ht]
\begin{center}
\mbox{\epsfig{figure=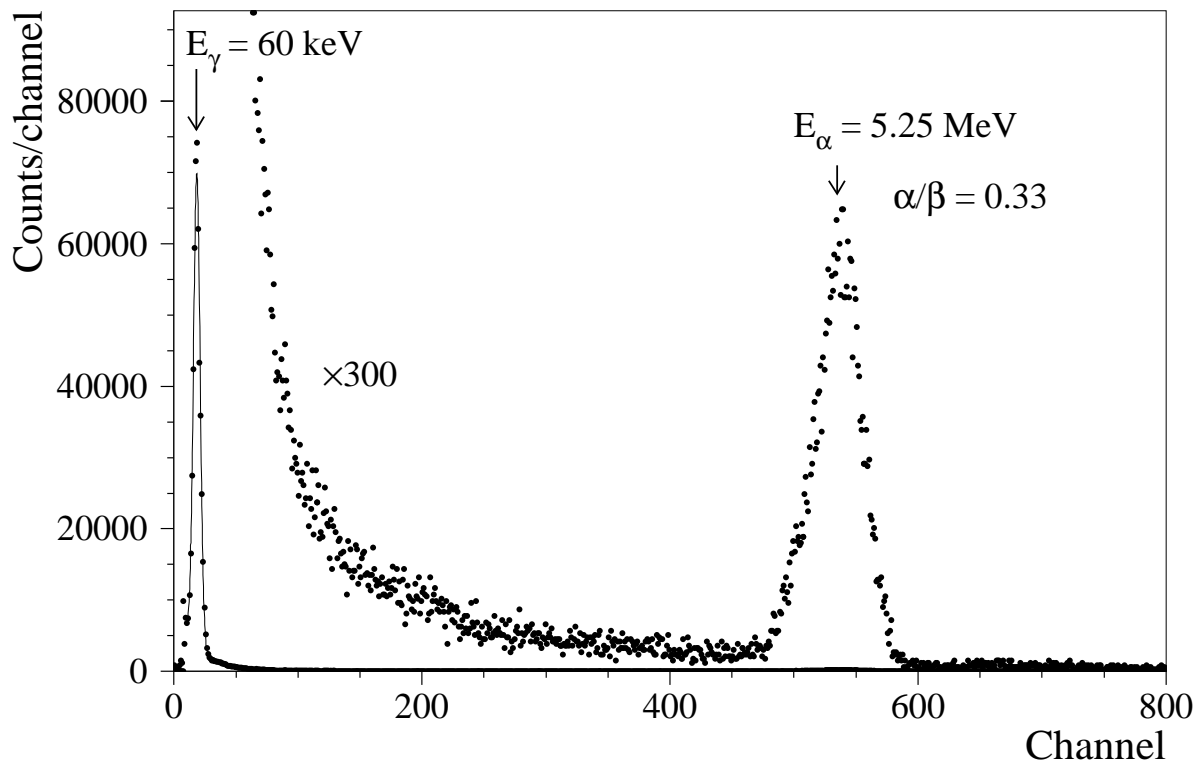,height=10.0cm}}
\caption {The energy spectrum of $\alpha$ particles from
collimated $^{241}$Am source.}
\end{center}
\end{figure}

\subsection{Pulse-shape discrimination for $\gamma$ quanta and $\alpha$
particles}

The shapes of scintillation light pulses in YAG:Nd crystals were
studied for $5.25$ MeV $\alpha$ particles and $\approx 1.5-1.8$
MeV $\gamma$ quanta with the help of a 12 bit 20 MHz transient
digitizer as described in \cite{W-alpha,Faz98}. The pulse shape
can be fitted by sum of exponential functions:
\begin{center}
$f(t)=\sum A_{i}/(\tau_{i}-\tau_{0})\cdot (e^{-t/\tau
_{i}}-e^{-t/\tau _{0}}),\qquad t>0$,
\end{center}
where $A_{i}$ are intensities (in \%), and $\tau_{i}$ are decay
constants for different light emission components, $\tau_{0}$ is
integration constant of electronics ($\approx$0.2~$\mu$s). The
values of $A_{i}$ and $\tau_{i}$ obtained by fitting the average
of $\approx$3 thousand individual $\alpha$ and $\gamma$ pulses
in the time interval $0-60$ $\mu$s (see Fig. 3) are presented in
Table 2. Observed difference in light pulse shapes allows to
discriminate $\gamma $($\beta$) events from $\alpha$ particles. We
applied for this purpose the optimal filter method proposed in
\cite{Gatti} and developed in \cite{Faz98}. To obtain the
numerical characteristic of YAG:Nd signal, so called shape
indicator ($SI$), the following formula was applied for each
pulse:

\begin{center}
$SI=\sum f(t_k)\cdot P(t_k)/\sum f(t_k)$,
\end{center}
where the sum is over time channels $k,$ starting from the origin
of pulse and up to 50 $\mu $s, $f(t_k)$ is the digitized amplitude
(at the time $t_k$) of a given signal. The weight function $P(t)$
is defined as: $P(t)=\{{f}_\alpha (t)-{f}_\gamma (t)\}/\{f_\alpha
(t)+f_\gamma (t)\}$, where $f_\alpha (t)$ and $f_\gamma (t)$ are
the reference pulse shapes for $\alpha$ particles and $\gamma$ quanta.
Reasonable discrimination between $\alpha$ particles and $\gamma$
rays was achieved using this approach as one can see in Fig. 4
where the shape indicator distributions measured by the YAG:Nd
scintillation crystal with $\alpha$ particles ($E_{\alpha}\approx
5.3$ MeV) and $\gamma$ quanta ($\approx 1.5-1.8$ MeV) are shown.

\nopagebreak
\begin{figure}[ht]
\begin{center}
\mbox{\epsfig{figure=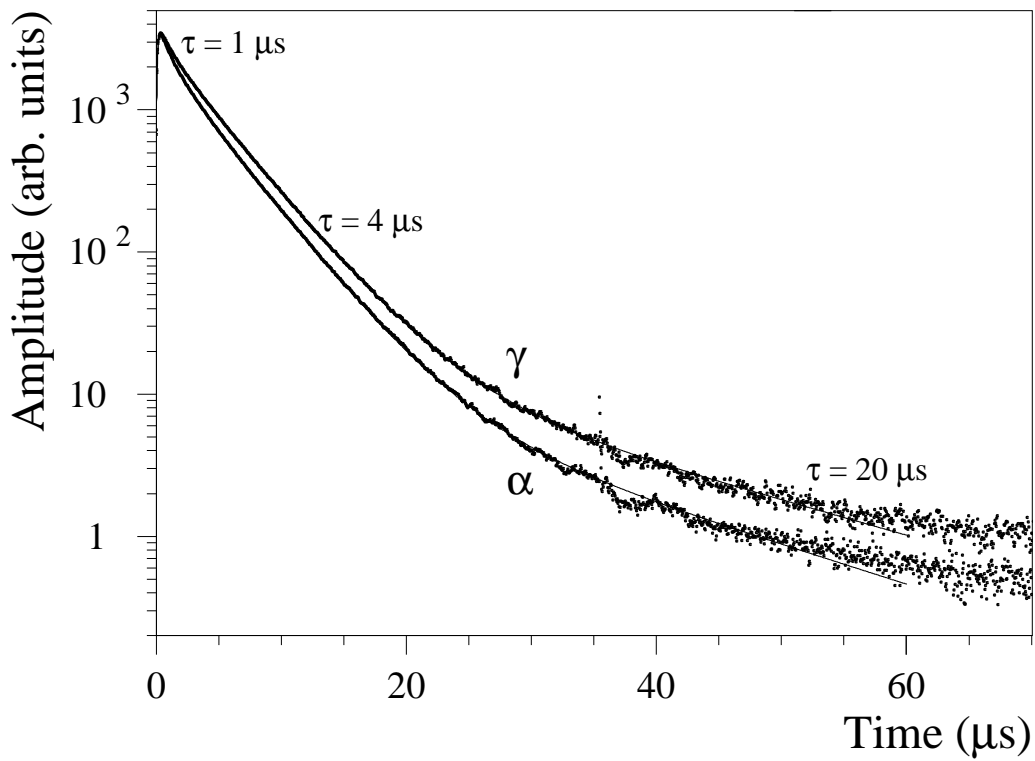,height=10.0cm}}
\caption {Decay of scintillation in YAG:Nd for $\gamma$ rays
and $\alpha$ particles ($\approx3000$ forms of $\gamma$ and
$\alpha$ events were added) and their fit by sum of three
exponential components with decay constants $\tau_i \approx 1$,
$\approx 4$, and $\approx 20$ $\mu$s.}
\end{center}
\end{figure}

\nopagebreak
\begin{figure}[ht]
\begin{center}
\mbox{\epsfig{figure=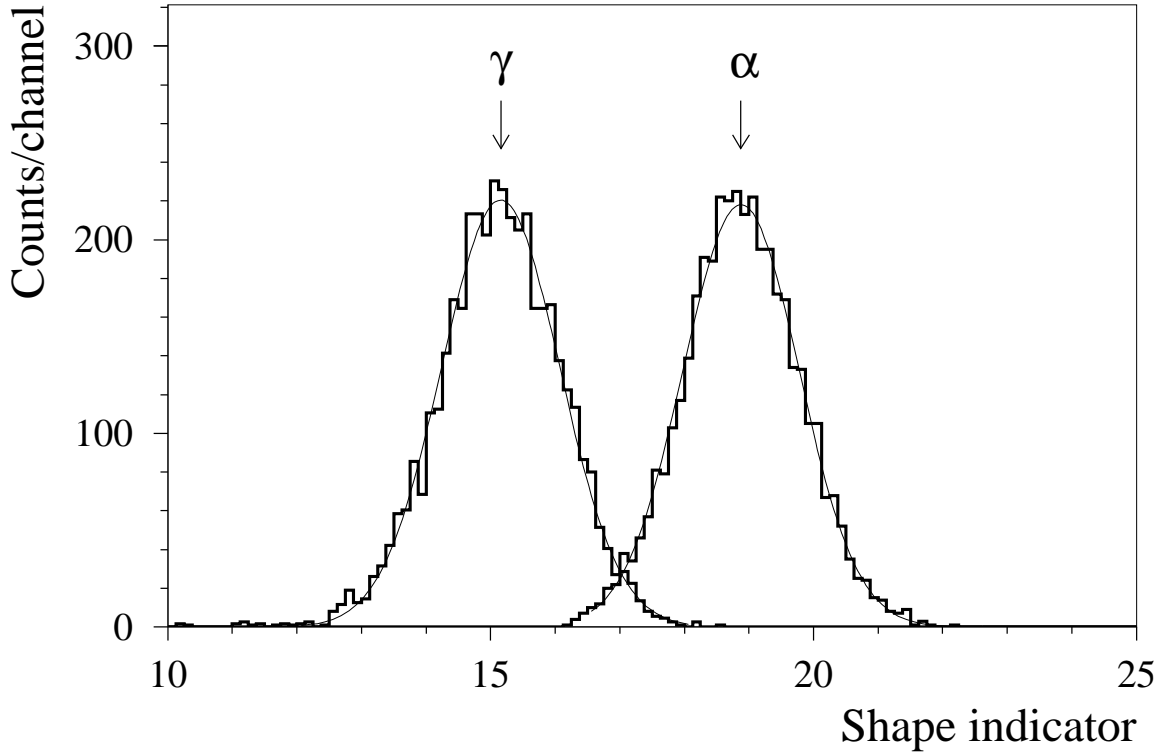,height=10.0cm}}
\caption {The shape indicator (see text) distributions
measured by YAG:Nd scintillation crystal with $\alpha$ particles
($E_{\alpha}=5.25$ MeV) and $\gamma$ quanta ($\approx 1.5-1.8$
MeV).}
\end{center}
\end{figure}

\begin{table}[tbp]
\caption{Decay time of YAG:Nd scintillator for $\gamma$ quanta and
$\alpha$ particles. The decay constants and intensities (in
percentage of the total intensity) are denoted as $\tau_i$ and
A$_i$, respectively.}
\begin{center}
\begin{tabular}{|l|l|l|l|l|}
\hline
Type of irradiation  &  \multicolumn{3}{|c|}{Decay constants, $\mu$s}  \\
\cline{2-4}
 ~   & $\tau_1$~(A$_1$)  & $\tau_2$~(A$_2$) & $\tau_3$~(A$_3$)\\
\hline
$\gamma$ rays & 1.1~(10\%) & 4.1~(87\%) & 20~(3\%)    \\
$\alpha$ particles & 1.0~(15\%) & 3.9~(83\%) & 18~(2\%)    \\
\hline
\end{tabular}
\end{center}
\end{table}

\subsubsection{Radioactive contamination of the YAG:Nd crystal}

To determine activity of $\alpha$ active nuclides from U/Th
contamination in the YAG:Nd crystal, the pulse shape analysis was
applied to data of 14.8 h low background measurement carried out
at the earth surface laboratory in Kiev. The YAG:Nd crystal was
viewed by the low-radioactive PMT (EMI D724KFLB) through plastic
scintillator light-guide 10 cm in diameter and 4 cm long. The
active light-guide reduces effect of $\gamma$ radiation from the
PMT and provides suppression of cosmic rays induced background due
to the pulse-shape discrimination of scintillation signals.
Passive shield was consisted of steel (2 cm) and lead (10 cm). The
energy resolution of the detector was 11\% for 662 keV $\gamma$
rays.

To estimate contamination of the crystal by $\alpha$ active
nuclides of U/Th families, the $\alpha$ events were selected from
the data accumulated with the YAG:Nd crystal with the help of the
pulse-shape discrimination technique. In the $\alpha$ spectrum
there is no peculiarities which can be surely attributed to
$\alpha$ active U/Th daughters. Analyzing these data we set the
limit on total $\alpha$ activity of $^{235}$U, $^{238}$U, and
$^{232}$Th daughters in the YAG:Nd crystal: $\leq 20$ mBq/kg.

\section{DISCUSSION}

\subsection{2$\beta$ decay of neodymium}

Among neodymium isotopes there are three possible $2\beta$
candidates: $^{146}$Nd, $^{148}$Nd, and $^{150}$Nd. They are
listed in Table 3 where the $Q_{2\beta}$ energies and the isotopic
abundances are given. As it was mentioned in the Introduction, the
most interesting of them is $^{150}$Nd due to the large value of
the transition energy. Two neutrino $2\beta$ decay of $^{150}$Nd
with the half-life $T_{1/2}=1.9^{+0.7}_{-0.4}\times10^{19}$ yr was
observed in experiment \cite{Art95} by using a time projection
chamber and samples of enriched $^{150}$Nd and natural neodymium.
In \cite{Sil97} the double beta decay of $^{150}$Nd was studied with
similar technique (only enriched $^{150}$Nd sample was used). The
two-neutrino half-life $T_{1/2}=(6.8\pm0.8)\times10^{18}$ yr was
measured.
Recently, the observation of $^{150}$Nd $2\nu 2\beta$ decay to the
second excited level of $^{150}$Sm ($0^+_1$ at 741 keV) with
$T_{1/2}=1.4^{+0.5}_{-0.4}\times10^{20}$ yr was reported in \cite{Bar04},
while only the limit $T_{1/2}>1.5\times10^{20}$ yr was determined
previously at 90\% C.L. \cite{Kli01}.
The most stringent limit on the neutrinoless mode
$T_{1/2}\geq1.2\times10^{21}$ yr was set in the experiment
\cite{Sil97}.

\begin{table}[tbp]
\caption{2$\beta$ unstable neodymium isotopes}
\begin{center}
\begin{tabular}{|c|c|c|}
\hline
 Transition & Isotopic abundance (\%) \cite{abundance} & $Q_{2\beta}$ (keV) \cite{Aud03} \\
 \hline
 $^{146}$Nd$ \rightarrow ^{146}$Sm & 17.2(0.3) & 70.2(2.9) \\
 $^{148}$Nd$ \rightarrow ^{148}$Sm & 5.7(0.1)  & 1928.8(1.9) \\
 $^{150}$Nd$ \rightarrow ^{150}$Sm & 5.6(0.2)  & 3367.5(2.2)  \\
 \hline
\end{tabular}
\end{center}
\end{table}

YAG:Nd scintillators provide a possibility to search for
$0\nu2\beta$ decay of $^{150}$Nd by "active" source experimental
method. The studied pulse-shape discrimination ability is an
important advantage of YAG:Nd scintillators as a low counting rate
detector. Background of scintillation detector caused by
radioactive contamination by U/Th daughters can be effectively
rejected due to pulse-shape analysis as it was demonstrated in
experiments with CdWO$_4$ \cite{W-alpha,Cd116} and CaWO$_4$
\cite{CARVEL,CaWO} crystal scintillators. Obvious disadvantage of
YAG:Nd detector is rather small concentration of Nd. In the
studied crystal the mass concentration is only $\approx$0.03\% of
$^{150}$Nd. It should be noted, however, that YAG:Nd crystals can be
produced with neodymium concentration up to 8 mol\%, which
corresponds to $\approx$0.1 mass\% of $^{150}$Nd. In this
connection we would like to refer the project CANDLES
\cite{CANDLES} which intends to use a few tons of non-enriched
CaF$_2$ crystals to search for $0\nu2\beta$ decay of $^{48}$Ca.
The concentration of target $^{48}$Ca nuclei in CaF$_2$ detector
is at the same level: $\approx$0.1 mass\% of $^{48}$Ca.

The value of energy resolution FWHM=4\% could allow $to$
$discover$ \cite{Zde04} the $0\nu2\beta$ decay of $^{150}$Nd with
the half-life $T_{1/2}^{0\nu2\beta}\sim 10^{25}$ yr, which
corresponds to the effective Majorana neutrino mass
$m_{\nu}\sim 0.06-0.6$ eV (taking into account all existing calculations of
nuclear matrix elements for $0\nu2\beta$ decay of $^{150}$Nd). As
for the $sensitivity$ \cite{Zde04},
an experiment involving $\approx$20~tons of
non-enriched YAG:Nd crystals with $\approx$8 mol\% of Nd could reach
the half-life sensitivity $\approx3\times10^{26}$ yr (supposing zero
background during ten years of measurements), which corresponds to
the Majorana neutrino mass $0.01-0.1$ eV.

\subsection{$\alpha$ decay of neodymium}

Alpha decay is allowed energetically for five naturally occurring
isotopes of neodymium. Their energies of $\alpha$ decay
($Q_{\alpha}$) and natural abundances are listed in Table 4. Calculation of
half-life values based on models \cite{Buc91,Poe83}, and the value
of the half-life of $^{145}$Nd calculated in \cite{Xu04} are also
presented. Alpha decay was observed only for $^{144}$Nd. Average
half-life $T_{1/2}=2.29\pm0.16\times10^{15}$ yr \cite{Son01} was
derived on the basis of four measurements:
$T_{1/2}=2.2\times10^{15}$ yr \cite{Por56},
$T_{1/2}=1.9\times10^{15}$ yr \cite{Iso65},
$T_{1/2}=2.4\pm0.3\times10^{15}$ yr \cite{Mac61} and
$T_{1/2}=2.65\pm0.37\times10^{15}$ yr \cite{Alb87}. It is obvious
that, due to low $Q_\alpha$ values
and long expected $T_{1/2}$'s, there is no perspective to
observe the $\alpha$ decay of $^{143}$Nd, $^{146}$Nd and
$^{148}$Nd. As for $^{145}$Nd, only two $T_{1/2}$ limits are
known: $T_{1/2}>6\times10^{16}$ yr \cite{Iso65} and
$T_{1/2}>1\times10^{17}$ yr \cite{Kau66}.

\begin{table}[tbp]
\caption{Theoretical calculations of half-lives for $\alpha$ decay
of natural Nd isotopes. Uncertainties given for calculations with
cluster model of Ref. \cite{Buc91} and semiempirical formula of Ref.
\cite{Poe83} are related only with uncertainty in the $Q_\alpha$
values. For $^{143}$Nd, we take into account that change in a
parity and spin additionally suppresses the $^{143}$Nd decay
rate.}
\begin{center}
\begin{tabular}{|lll|ccc|c|}
 \hline
 Isotope    & Abund.           & $Q_\alpha$, MeV & \multicolumn{3}{|c|}{Calculated $T_{1/2}$, yr} & Exp. $T_{1/2}$, yr              \\
 ~          & \cite{abundance} & \cite{Aud03}    & based on \cite{Buc91}             & based on \cite{Poe83}            & \cite{Xu04}       & ~ \\
 \hline
 ~ & ~ & ~ & ~ & ~ & ~ & ~\\
 $^{143}$Nd & 12.2\%       & 0.520(7)        & $>1.1_{-1.0}^{+7.8}$$\times$$10^{80}$ & $5.2_{-4.7}^{+51}$$\times$$10^{93}$  & --                 & -- \\
 $^{144}$Nd & 23.8\%       & 1.905(2)        & $1.9_{-0.1}^{+0.2}$$\times$$10^{15}$  & $4.3_{-0.3}^{+0.3}$$\times$$10^{15}$ & --                 & $2.29\pm0.16$$\times$$10^{15}$\\
 $^{145}$Nd & 8.3\%        & 1.578(2)        & $1.7_{-0.1}^{+0.2}$$\times$$10^{22}$  & $3.9_{-0.4}^{+0.3}$$\times$$10^{23}$ & $3.7$$\times$$10^{22}$ & $>1$$\times$$10^{17}$\\
 $^{146}$Nd & 17.2\%       & 1.182(2)        & $2.0_{-0.4}^{+0.4}$$\times$$10^{34}$  & $3.9_{-0.7}^{+0.8}$$\times$$10^{34}$ & --                 & --\\
 $^{148}$Nd & 5.7\%        & 0.599(3)        & $6.1_{-3.1}^{+6.7}$$\times$$10^{70}$  & $1.1_{-0.6}^{+1.2}$$\times$$10^{71}$ & --                 & --\\
 ~ & ~ & ~ & ~ & ~ & ~ &~\\ \hline
\end{tabular}
\end{center}
\end{table}

The indication on the $\alpha $ decay of natural $^{180}$W isotope
with half-life $T_{1/2}=1.2_{-0.4}^{+0.8}$ (stat) $\pm 0.3$ (syst)
$\times 10^{18}$ yr has been recently observed in the experiment
\cite{W-alpha} with the help of the low background $^{116}
$CdWO$_4$ crystal scintillators. This result was confirmed in the
measurements with CaWO$_4$ crystal as scintillator \cite{CaWO} and
bolometer \cite{Coz04}. It should be also referred the excellent
result on the detection of alpha decay of $^{209}$Bi with the
half-life $T_{1/2}=(1.9\pm 0.2)\times 10^{19}$ y
\cite{Mar03}. In both the experiments \cite{Coz04} and
\cite{Mar03} the cryogenic technique, which uses simultaneous
registration of heat and light signals, have been applied. This
method provides perfect selection of $\alpha$ events on the
background caused by $\gamma$ rays (electrons). It should be
interesting to check YAG:Nd crystal as the cryogenic detector to
search for alpha decay of $^{145}$Nd and to measure more accurately
the $^{144}$Nd half-life.

It should be also mentioned a possibility to search for
spin-dependent inelastic scattering of weakly interacting massive
particles (WIMP) with excitation of low energy nuclear levels of
$^{145}$Nd (the lowest one: $3/2^-$ 67 keV, E2 transition, and the
second level: $5/2^-$ 72 keV, M1 transition) due to nonzero spin
of this nucleus ($7/2^-$). Identification of such "mixed" (nuclear
recoil plus $\gamma$ quanta) events could be possible due to the
simultaneous registration of heat and light signals (such a
techniques have been used in \cite{CRESST,ROSEBUD}). A heat/light
ratio for such events would differ from "pure" nuclear recoils or
$\gamma$(electron) events.

\section{CONCLUSIONS}

The scintillation properties of YAG:Nd crystals were studied. The
energy resolution 9.3\% (662 keV $^{137}$Cs $\gamma$ line) was
obtained with the YAG:Nd crystal scintillator placed in liquid and
viewed by two PMTs. Shapes of scintillation signals were
investigated, and reasonable pulse-shape discrimination for
$\gamma$ rays and $\alpha$ particles was achieved. The
$\alpha/\beta$ ratio was measured with the YAG:Nd scintillator to
be equal 0.33 at the energy of alpha particles 5.25 MeV.
Radioactive contamination of the YAG:Nd crystal by $\alpha$ active
nuclides from U/Th chains was estimated as $\leq20$ mBq/kg.

Three potentially $2\beta$ active neodymium isotopes can be studied
with the help of YAG:Nd crystal. Due to good scintillation
characteristics and pulse-shape discrimination ability the YAG:Nd
scintillators seems to be encouraging material for the $2\beta$
experiment with $^{150}$Nd.

The YAG:Nd crystals could be used to search for  $\alpha$ decay of
natural neodymium isotopes, in particular, $^{145}$Nd. A strong
signature of $\alpha$ decay can be assured via simultaneous
registration of heat and light signals using cryogenic technique.
However, ability of YAG:Nd crystal as cryogenic detector has to
be elaborated in an experiment.

\section{ACKNOWLEDGEMENT}

The authors would like to thank Dr. Yu.D.~Glinka from the Institute of
Surface Chemistry of the National Academy of Sciences of Ukraine
for lending the YAG:Nd crystal used in the present study.

\end{document}